\documentclass[reprint,10pt,5p,times,sort&compress,twocolumn]{elsarticle}
\usepackage{amsfonts}
\usepackage{amsmath} 
\usepackage{amssymb} 
\usepackage{graphicx}
\usepackage{color}
\journal{ Communications in Nonlinear Science and Numerical Simulation }
\begin{document}
\begin{frontmatter}
\title{Dust-acoustic solitary waves in a magnetized dusty plasma with nonthermal  electrons and trapped ions}
\author[APM]{A. P. Misra\corref{cor1}}
\ead{apmisra@visva-bharati.ac.in, apmisra@gmail.com}
\cortext[cor1]{Corresponding author.}
\address[APM]{Department of Mathematics, Siksha Bhavana, Visva-Bharati University, Santiniketan-731 235, West Bengal, India}
\author[YW]{Yunliang Wang}
\ead{ylwang@ustb.edu.cn}
\address[YW]{Department of Physics, School of Mathematics and Physics,
University of Science and Technology Beijing, Beijing 100083, China}

\begin{abstract}
The   nonlinear propagation of electrostatic dust-acoustic (DA)  waves in a magnetized   dusty plasma   consisting of negatively  charged mobile dusts, nonthermal fast electrons and trapped ions with vortex-like distribution is studied.  Using the  reductive perturbation technique,  a Korteweg-de Vries (KdV)-like equation is derived which governs the dynamics of the small-amplitude solitary waves in a   magnetized dusty nonthermal plasma. It is found that due to the dust thermal pressure, there exists a critical value $(\beta_c)$ of the nothermal parameter $\beta~(>1)$, denoting the percentage of energetic electrons, below which the  DA solitary waves cease  to propagate.  The  soliton solution (travelling wave) of the KdV-like equation is obtained, and is shown to  be only of the rarefactive type. The   properties of the solitons are analyzed numerically with the system parameters.  It is also seen that the effect of the static magnetic field (which only modifies the soliton width) becomes significant when the dust gyrofrequency is smaller than one-tenth of  the dust plasma frequency. Furthermore, the amplitude of the soliton is found to increase (decrease) when the   ratio of the free to trapped ion temperatures $(\sigma)$ is positive  (negative).
The effects of the system parameters including the obliqueness of  propagation $(l_z)$ and $\sigma$ on the dynamics of the DA solitons are also discussed numerically, and it is found that the soliton structures can withstand perturbations and turbulence during a considerable time. The results should be useful for understanding the nonlinear propagation of DA solitary waves  in  laboratory and space plasmas (e.g., Earth's magnetosphere, auroral region, heliospheric environments etc.).

\end{abstract}
\end{frontmatter}
\section{Introduction}
Since the work of Rao {\it et al.} \cite{rao1990}, in which the existence of nonlinear DA waves in unmagnetized dusty plasmas (where the charged dusts provide the inertia and the restoring forces come from the pressures of inertialess electrons and ions) was  first predicted, there has been a number of works   focusing on the linear or nonlinear properties of   low-frequency DA waves  in magnetized or unmagnetized plasmas (See, e.g.,  Refs. \cite{shukla2001,shukla2009,pintu2008,animesh2012,deka2012,wang2009,wang2010,misra2013,merlino1998}). The nonlinear propagation of such waves can give rise to the formation of solitons with negative or positive wave amplitudes, which has
 potential applications in astrophysical and space environments (such as cometary tails, planetary rings, interstellar clouds and lower parts of Earth's ionosphere etc.) as well as in laboratory and technological studies \cite{mendis1994,verheest1996,verheest2000,shukla2002}. Furthermore,  electrostatic solitary waves   have been observed in several regions, including the Earth’s magnetotail, bow shock/solar wind, and polar magnetosphere \cite{franz1998,pickett2003}.

On the other hand, the observations in space plasma environments indicate the presence of electrons and ions which are not in thermodynamic equilibrium \cite{lundin1989,futaana2003}.   The  presence of nonthermal electrons in Earth's bow-shock region has   been confirmed  by the  Vela satellite \cite{lundin1989}. Also, the ASPERA on the Phobos 2 satellite has recorded the loss of energetic electrons from the upper ionosphere of Mars.    Furthermore,  observations made by the Viking spacecraft  and Freja satellite  have indicated the presence of  electrostatic solitary structures  in the magnetosphere with density depressions \cite{dovner1994}. It has been shown by  numerical simulation that  the propagation of DA waves can give rise to a significant amount of ion trapping in the wave potential \cite{winske1995}. Malkki {\it et al.} \cite{malkki1989} had demonstrated that the nonlinear ion-hole instability is the best candidate for the generation mechanism of the negative isolated potentials.
  Naturally, there is a departure from the Boltzmann  distribution and one thus  encounters a vortex-like ion distribution  in phase space. The nonthermal distribution of electrons with an excess of energetic particles can be modelled by the Cairns distribution \cite{cairns1995,cairns1996}, while the distribution of free as well as trapped particles can be considered as prescribed by Schamel  \cite{schamel1972}.

The important effects of the external magnetic field \cite{das2001,lin2010,zhang2010}  as well as the nonthermal electrons  \cite{mamun1998a,bandyopadhyay1999,misra2002,moslem2005a,moslem2005b,paul2013} and trapped ions with vortex-like distribution \cite{zahran2013,duan2003,mamun1998b,paul2013} on the properties of electrostatic DA waves have been studied in magnetized or unmagnetized plasmas. Some analytical studies on the nonlinear propagation of DA waves in   unmagnetized plasmas \cite{lin2007} as well as plasmas with vortex-like ions \cite{lin2007a} have also been reported. Using variable separation approach, Lin {\it et al.} \cite{lin2007b} had investigated the nonlinear features of DA solitary waves with dust charge fluctuation.  Furthermore, the properties of solitary waves and shocks (SWS) have been studied in other environments, e.g., dense quantum plasmas to show that quantum parameter plays an important role in the formation of compressive or rarefactive SWS \cite{bagchi2010}.     Recently, it has been shown that the dusty plasmas consisting of vortex-like ions and nonisothermal electrons, both for the first- and second-order perturbed potentials,
support only rarefactive solitons. It was pointed out that    the higher-order approximation enhances the amplitude of the DA waves  \cite{eslami2013}. Furthermore, the nonlinear propagation of  DA waves has been investigated in unmagnetized dusty plasmas consisting of Boltzmannian \cite{mamun1998b} or nonthermal \cite{paul2013} electrons and vortex-like ions . However, no theory has been reported so far for the oblique propagation of DA waves in a magnetized dusty plasma with nonthermal electrons and trapped ions.

In this paper, we investigate the nonlinear propagation of electrostatic DA waves in a magnetized   dusty plasma consisting of   negatively   charged dust fluid, non-thermal fast electrons and trapped ions with vortex-like distribution. We use the reductive perturbation technique (RPT) to derive a KdV-like equation which has nonlinearity proportional to three-half power of the wave potential. It is shown that solitary waves may propagate as stationay solitons of the rarefactive type. The properties of the soliton as well as its dynamical evolution are analyzed with the system parameters. The results should be useful for understanding the nonlinear features of localized electrostatic perturbations in laboratory and space plasmas (e.g., Earth's magnetosphere, auroral region, heliospheric environments) in which magnetized negatively charged dusts, nonthermal electrons and trapped ions are the major plasma species.

\section{Basic equations and the derivation of KdV-like equation}

We consider the nonlinear propagation of DA waves in  a magnetized dusty plasma consisting of very massive, micrometer-sized, negatively charged inertial dust grains, nonthermally distributed fast electrons, and ions with trapped particles.   The external static magnetic field is considered along the $z$-axis, i.e.,  ${\bf B}=B_{0} \hat{\bf z}$. Thus, the nonlinear dynamics of the low-phase velocity (in comparison with the electron and ion thermal velocities, i.e., $v_{td}\ll\omega/k\ll v_{te},v_{ti}$, where $v_{tj}=\sqrt{k_BT_j/m_j}$ is the thermal velocity of $j$-species particles with $j=e,~i,~d$ for electrons, ions and charged dusts), low-frequency (in comparison with the dust-cyclotron frequency) DA waves is given by the following set of   normalized equations:
\begin{equation}
\frac{\partial n_{d}}{\partial t}+\nabla \cdot \left(n_{d} {\bf u}_{d}\right)=0, \label{cont-eq}
\end{equation}
\begin{eqnarray}
\left(\frac{\partial }{\partial t}+ {\bf u}_{d} \cdot \nabla \right) {\bf u}_{d}&& = \nabla \phi
   +\omega_{c}\left({\bf u}_{d}\times \hat{\bf z}\right)\notag\\
   &&-\frac{5}{3}T n_{d}^{-{1}/{3}}\nabla n_{d}, \label{moment-eq}
\end{eqnarray}
\begin{equation}
 \nabla^{2}\phi=\delta n_{e}-\mu n_{i}+n_{d}, \label{poisson-eq}
\end{equation}
where $n_{d}$ and ${\bf u}_d$ are, respectively, the   number density and velocity of charged dusts (with mass $m_d$) normalized by the unperturbed value $n_{d0}$ and the DA speed $c_{d}=\sqrt{Z_{d}k_{B}T_{e}/m_{d}}$, with $T_{e}$ denoting the electron temperature, $k_B$ the Boltzmann constant and $Z_{d}$    the charged dust state, i,e., the number of electrons/ions residing on the dust-grain surface. Also, $\phi$ is the electrostatic wave potential normalized by $k_BT_{e}/e$, where $e$ is the magnitude of the electron charge, $T=T_{d}/T_{e}$ with $T_{d}$ denoting the dust temperature, $\omega_{c}=|q_{d}| B_{0} / m_{d} \omega_{pd}$ is the dust-cyclotron frequency normalized by the dust plasma oscillation frequency $\omega_{pd}=\sqrt{4\pi n_{d0} Z_{d}^{2}e^2/m_{d}}$. The time and space variables are, respectively, in the units of the dust plasma period $\omega^{-1}_{pd}$  and the Debye length $\lambda_{D}=\sqrt{k_{B} T_{e} / 4 \pi   n_{d0}Z_{d}e^{2}}$. At equilibrium,   the  charge neutrality condition reads $\mu\equiv n_{i0} / Z_{d} n_{d0}=1+\delta\equiv1+ n_{e0} / Z_{d} n_{d0}$. For the propagation of DA waves since the thermal motion of charged dusts can not keep up with the wave, we consider adiabatic compression of the dust fluid, and use the pressure law: $P_d/P_{d0}=(n_d/n_{d0})^{\gamma}$, where $P_{d0}=k_BT_dn_{d0}$ and $\gamma=5/3$ for three-dimensional configuration. In this assumption,  the effects
of viscosity, thermal conductivity and the energy transfer due to collisions can   be neglected. In our fluid model  we have also considered the plasma thermal pressure to be smaller than the magnetic pressure and the charging of the dust grains is held constant.

In what follows,  the nonthermal distribution of electrons (fast or energetic particles) are given by \cite{cairns1995,cairns1996}
\begin{equation}
  n_{e}=(1-\beta\phi+\beta\phi^{2})e^{\phi} \label{elec-dens}.
\end{equation}
where $\beta=4\gamma/1+3\gamma$, with $\gamma>0$,  denotes the degree of nonthermality of the charged particles (percentage of fast or energetic electrons) in the plasma, and $\beta\gtrless1$ according to when $\gamma\gtrless1$. The value $\gamma=0$ corresponds to the case of thermal equilibrium (Boltzmann distribution) of electrons.

On the other hand, the ion distribution with free and trapped particles is modelled  for small-amplitude perturbations ($|\phi|\ll1$) as \cite{schamel1972}
\begin{equation}
   n_{i}\approx1-\phi-\frac{4(1-\sigma)}{3\sqrt{\pi}}(-\phi)^{{3}/{2}}+\frac{1}{2}\phi^{2}, \label{ion-dens}
\end{equation}
Note that this expression for $n_i$ (which becomes the same for $\sigma<0$ and $\sigma>0$ in the small-amplitude limit)  is obtained after integrating the   ion distribution functions over the velocity space $(v)$ for free $(f_{if})$ and trapped $(f_{it})$ particles which solve the corresponding Vlasov equation. The distribution functions are \cite{schamel1972}
\begin{eqnarray}
\begin{split}
f_{if}=\frac{1}{\sqrt{2\pi}}\exp\left[-\frac{1}{2}\left(v^2+2\phi\right)\right],~|v|>\sqrt{-2\phi},  \\
f_{it}=\frac{1}{\sqrt{2\pi}}\exp\left[-\frac{\sigma}{2}\left(v^2+2\phi\right)\right],~|v|\leq\sqrt{-2\phi},
\end{split}
\end{eqnarray}
where $\sigma=T_{if}/T_{it}$ is the ratio of  the free to trapped ion temperatures and determines the densities of trapped ions. From a qualitative plot of these distribution functions, it can be seen that the negative values $(\sigma<0)$   of the temperature ratio $\sigma$   correspond  to an excavated vortex particle distribution, i.e.,  a depression in the trapped particle distribution, while $\sigma=0$ represents the plateau (flat-topped). It has been shown that the negative values of the trapping parameter $\sigma$ may lead to the generation of ion holes \cite{schamel2008} and solitary phase-space holes \cite{eliasson2005} in plasmas. Furthermore,  in the limit of $\sigma\rightarrow1$, $n_i(\phi)$ approaches the Boltzmann distribution.  Thus, one can consider either $\sigma<0$ or $0<\sigma<1$.  However, as  shown in Ref. \citep{schamel1972} that the wave steepening parameter is proportional to $1-\sigma$. So, when $\sigma>1$, i.e., the influence of the trapped ions is inverted or when the number of trapped ions is increased, there may exist a   more steepened wave, and because of the more peaked bump of the ion distribution, a stability analysis would exhibit the unstable nature of these waves. Thus, waves with $\sigma>1$ may be physically unrealistic.  

Next, for the dynamical evolution of the small-amplitude electrostatic DA perturbations we   follow the same RPT as in Refs. \cite{mamun1998a,mamun1998b,paul2013}.  Note that the reductive perturbation method is a very powerful way of deriving nonlinear evolution equations in simplified forms for the propagation of different kind of waves and their interactions in various nonlinear media. The method has   been well applied not in plasma or fluid media, but also in other nonlinear media, e.g., in a liquid with gas bubbles to study the characteristic features of nonlinear waves  \cite{kudryashov2013,kudryashov2014}.  For the description of RPT and their applications readers are referred to the work of Leblond  \cite{leblond2008}.    Thus,  the independent variables are stretched as
\begin{eqnarray}
&&\xi=\epsilon^{{1}/{4}}(l_{x}x+l_{y}y+l_{z}z-v_{0}t), \notag\\
&&\tau=\epsilon^{{3}/{4}}t, \label{stretch-coord}
\end{eqnarray}
where $\epsilon~(0<\epsilon<1)$ is a small  parameter measuring the weakness of the wave amplitude, $v_{0}$ is the nonlinear wave phase velocity normalized by the DA speed $c_{d}$. Also, $l_x,~l_y$ and $l_z$ are the direction cosines of the wave vector along the axes such  that $l_x^2+l_y^2+l_z^2=1$.  The dependent variables, namely $n_{d},u_{d}$ and $\phi$ are expanded as
\begin{equation}
\begin{split}
&n_{d}=1+\epsilon n_{d}^{(1)}+\epsilon^{{3}/{2}} n_{d}^{(2)}+ \cdots, \\
&\phi=\epsilon\phi^{(1)}+\epsilon^{{3}/{2}}\phi^{(2)}+\cdots,\\
& u_{dz}=\epsilon u_{z}^{(1)}+\epsilon^{{3}/{2}} u_{z}^{(2)}+\cdots,\\
& u_{d(x,y)}=\epsilon^{{5}/{4}} u_{x,y}^{(1)}+\epsilon^{{3}/{2}} u_{x,y}^{(2)}+\cdots. \label{perturb-quant}\\
\end{split}
\end{equation}
In the above expansions,   the first-order perturbations for the transverse velocity components of the dust fluids appear in higher-orders of  $\epsilon$ than that for the
parallel component. For the nonlinear DA waves, this anisotropy is introduced due to the fact that the dust gyromotion (perpendicular to the magnetic field) is treated as a higher-order effect than the motion parallel to the magnetic field \cite{misra2006}.  

In what follows, we substitute Eqs. \eqref{stretch-coord} and \eqref{perturb-quant}  into Eqs. \eqref{cont-eq}-\eqref{poisson-eq}, and equate the coefficients of different powers of $\epsilon$. Thus, equating the coefficients of $\epsilon^{{5}/{4}}$, $\epsilon^{{6}/{4}}$ and   $\epsilon^{{7}/{4}}$, from Eq. \eqref{cont-eq}, we successively obtain
\begin{equation}
-v_{0}n_{d}^{(1)}+l_{z}u_{z}^{(1)}=0, \label{cont-1}
\end{equation}
\begin{equation}
l_{x}u_{x}^{(1)}+l_{y}u_{y}^{(1)}=0, \label{cont-2}
\end{equation}
\begin{equation}
-v_{0}\frac{\partial n_{d}^{(2)}}{\partial\xi}+\frac{\partial n_{d}^{(1)}}{\partial\tau}+l_{x}\frac{\partial u_{x}^{(2)}}{\partial\xi}+l_{y}\frac{\partial u_{y}^{(2)}}{\partial\xi}+l_{z}\frac{\partial u_{z}^{(2)}}{\partial\xi}=0. \label{cont-3}
\end{equation}
From the $x$-component of Eq. \eqref{moment-eq}, equating the coefficients   of $\epsilon^{{5}/{4}}$ and $\epsilon^{{6}/{4}}$ we obtain
\begin{equation}
l_{x}\frac{\partial\phi^{(1)}}{\partial\xi}+\omega_{c}u_{y}^{(1)}-\frac{5}{3}T l_{x}\frac{\partial n_{d}^{(1)}}{\partial\xi}=0, \label{moment-x1}
\end{equation}
\begin{equation}
-v_{0}\frac{\partial u_{x}^{(1)}}{\partial\xi}=\omega_{c}u_{y}^{(2)}. \label{moment-x2}
\end{equation}
Also, from the $y$-component of Eq. \eqref{moment-eq}, equating the coefficients   of $\epsilon^{{5}/{4}}$ and $\epsilon^{{6}/{4}}$ we obtain
\begin{equation}
l_{y}\frac{\partial\phi^{(1)}}{\partial\xi}-\omega_{c}u_{x}^{(1)}-\frac{5}{3}T l_{y}\frac{\partial n_{d}^{(1)}}{\partial\xi}=0, \label{moment-y1}
\end{equation}
\begin{equation}
v_{0}\frac{\partial u_{y}^{(1)}}{\partial\xi}=\omega_{c}u_{x}^{(2)}. \label{moment-y2}
\end{equation}
Similarly, from the $z$-component of Eq. \eqref{moment-eq}, equating the coefficients   of $\epsilon^{{5}/{4}}$ and $\epsilon^{{7}/{4}}$ we successively obtain
\begin{equation}
v_{0}u_{z}^{(1)}=-l_{z}\phi^{(1)}+\frac{5}{3}Tl_{z}n_{d}^{(1)}, \label{moment-z1}
\end{equation}
\begin{equation}
-v_{0}\frac{\partial u_{z}^{(2)}}{\partial\xi}+\frac{\partial u_{z}^{(1)}}{\partial\tau}= l_{z}\frac{\partial\phi^{(2)}}{\partial\xi}-\frac{5}{3}T l_{z}\frac{\partial n_{d}^{(2)}}{\partial\xi}. \label{moment-z2}
\end{equation}
Next, using the expressions \eqref{elec-dens} and \eqref{ion-dens} for electron  and ion number densities, we successively obtain after equating the coefficients of $\epsilon$ and $\epsilon^{3/2}$ from  Eq. \eqref{poisson-eq} as
\begin{equation}
n_{d}^{(1)}=-\left[\mu+\delta(1-\beta)\right]\phi^{(1)}, \label{poisson-1}
\end{equation}
\begin{eqnarray}
n_{d}^{(2)}&&=\frac{\partial^{2}\phi^{(1)}}{\partial\xi^{2}}-\frac{4\mu(1-\sigma)}{3\sqrt{\pi}}\left(-\phi^{(1)}\right)^{{3}/{2}}\notag \\
&&-[\mu+\delta(1-\beta)]\phi^{(2)}.\label{poisson-2}
\end{eqnarray}
Using Eq. \eqref{poisson-1}, we obtain from Eqs. \eqref{moment-x1}, \eqref{moment-y1} and \eqref{moment-z1} the first-order components of the dust fluid velocity as
\begin{equation}
u_{x}^{(1)}=\frac{l_{y}}{\omega_{c}}\left[1+\frac{5}{3}T(\mu+\delta(1-\beta))\right]\frac{\partial\phi^{(1)}}{\partial\xi}, \label{ux1}
\end{equation}
\begin{equation}
u_{y}^{(1)}=-\frac{l_{x}}{\omega_{c}}\left[1+\frac{5}{3}T(\mu+\delta(1-\beta))\right]\frac{\partial\phi^{(1)}}{\partial\xi},\label{uy1}
\end{equation}
\begin{equation}
u_{z}^{(1)}=-\frac{v_{0}}{l_{z}}\left[\mu+\delta(1-\beta)\right]\phi^{(1)}. \label{uz1}
\end{equation}
Similarly, using Eqs. \eqref{ux1} and \eqref{uy1},   we obtain from Eqs. \eqref{moment-x2} and \eqref{moment-y2} the second order quantities  for the velocity as
\begin{equation}
u_{x}^{(2)}=-\frac{v_{0}l_{x}}{\omega_{c}^{2}}\left[1+\frac{5}{3}T(\mu+\delta(1-\beta))\right]\frac{\partial^{2}\phi^{(1)}}{\partial\xi^{2}},\label{ux2}
\end{equation}
\begin{equation}
u_{y}^{(2)}=-\frac{v_{0}l_{y}}{\omega_{c}^{2}}\left[1+\frac{5}{3}T(\mu+\delta(1-\beta))\right]\frac{\partial^{2}\phi^{(1)}}{\partial\xi^{2}}. \label{uy2}
\end{equation}
Eliminating $u_{z}^{(1)}$ and $n_{d}^{(1)}$ from Eqs. \eqref{moment-z1}, \eqref{poisson-1}  and \eqref{uz1}  we obtain the following dispersion relation
\begin{equation}
v_{0}=l_{z}\left[\frac{1}{\mu+\delta(1-\beta)}+\frac{5}{3}T\right]^{1/2}. \label{disp-rel}
\end{equation}
This represents the phase velocity of the obliquely propagating DA waves in the moving $\xi-\tau$ frame of reference. It can be shown  that this phase velocity corresponds to the longitudinal propagation of low-frequency, long-wavelength slow-wave eigen modes that can be obtained by Fourier analyzing the set of basic  equations. Furthermore, for the DA waves to propagate with the velocity $v_0$, the expression in the square brackets of Eq. \eqref{disp-rel} must be positive, which gives  $\beta>\beta_c\equiv1+(\mu+3/5T)/\delta>1$. This implies that in hot dusty plasmas, the small-amplitude DA wave propagation is possible when the percentage of energetic electrons (or the degree of nonthermality) exceeds some critical value and  is higher than the nonenergetic ones. However, in cold dusty plasmas ($T=0$), the same can propagate with $\beta<1$ irrespective of the polarity (positive or negative) of the charged dusts. Comparing    Eq. \eqref{disp-rel} with Eq. (16) in Ref. \cite{paul2013}, we find that the phase velocity $v_0$ is not only influenced by the effects of the number density of charged particles and the nonthermality of electrons, but also greatly modified by  the effects of the dust temperature $(T_d)$ as well as  the obliqueness of propagation ($l_z$).

\begin{figure*}[ht]
\centering
\includegraphics[height=2.5in,width=6in]{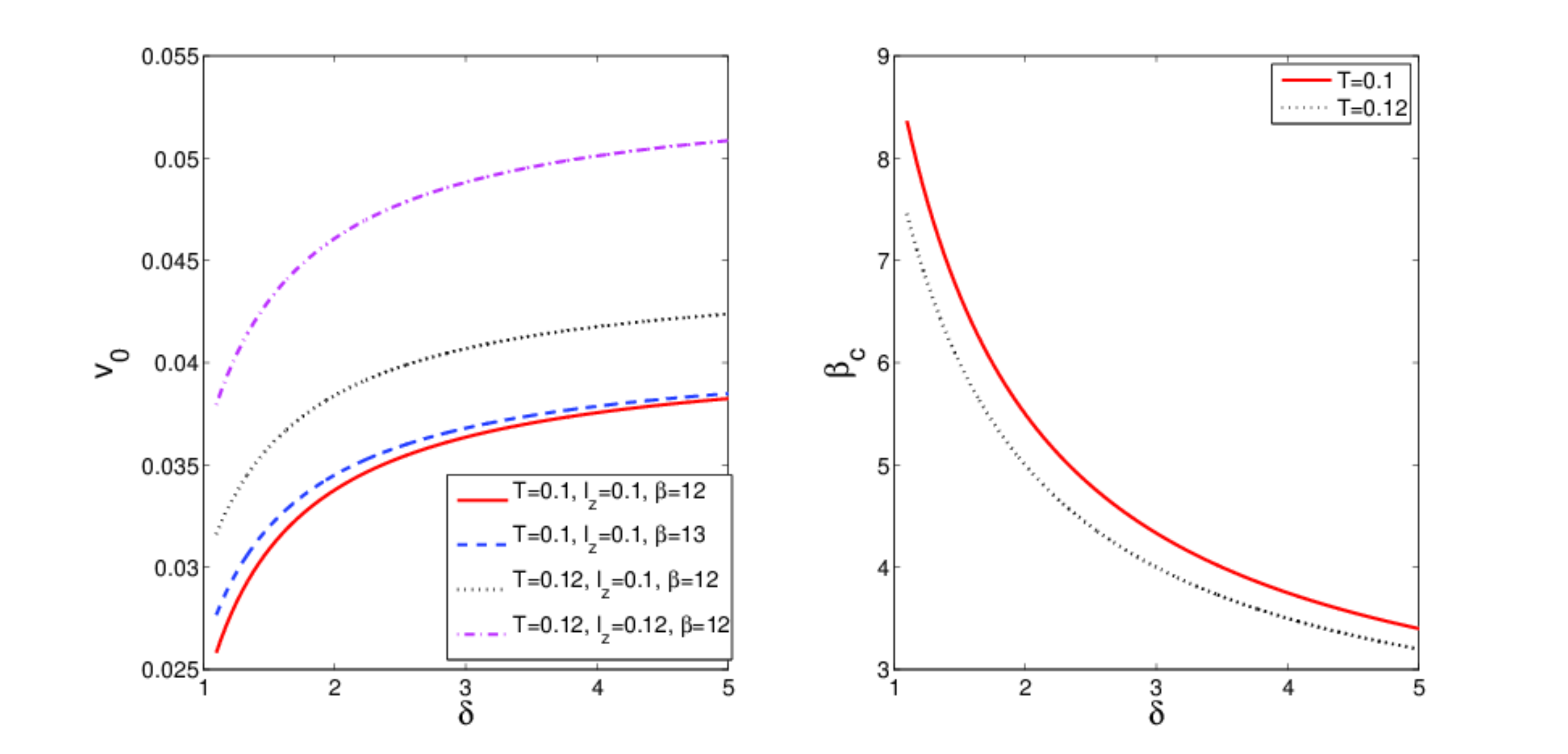}
\caption{The left panel shows the typical variations of the phase velocity $v_0$ [Eq. \eqref{disp-rel}] with the density ratio $\delta$  for different values of $\beta$, $T$ and $l_z$ as shown in the figure. The right panel is a plot of $\beta_c$ (critical value of $\beta$ for which $v_0$ is real, i.e., DA wave propagation is possible) with   $\delta$   for different values of the temperature ratio $T$ as in the figure.}
\label{fig:figure1}
\end{figure*}

 Typical variations  of the phase velocity $v_0$ with respect to the density ratio $\delta$ are shown in Fig. \ref{fig:figure1} (See the left panel)   for different values of $l_z$, $T$ and $\beta$. We find that the phase velocity of the electrostatic perturbations is always less than the DA speed $c_d$, i.e., $v_{td}\ll v_0<c_d\ll v_{ti}<v_{te}$ is satisfied.   It approaches a constant value at higher values of $\delta$.    Furthermore,  the values of  $v_0$  increases (decreases) with increasing (decreasing) values of the parameters $l_z$ (obliqueness of the angle of propagation), the nonthermal parameter $\beta~(>\beta_c)$ and the temperature ratio $T~(\propto T_d)$. From the right panel of Fig. \ref{fig:figure1}, which shows the variation of $\beta_c$ with respect to $\delta$, one can conclude that the larger the values of the dust temperature, the smaller is the percentage of fast electrons to be required to excite the electrostatic DA waves.

Next, from Eqs. \eqref{cont-3} and \eqref{moment-z2}   we obtain
\begin{eqnarray}
\left(v_{0}-\frac{5}{3}\frac{l_{z}^{2}}{v_{0}}T\right)\frac{\partial n_{d}^{(2)}}{\partial\xi}&&=\frac{\partial n_{d}^{(1)}}{\partial\tau}-\frac{l_{z}^{2}}{v_{0}}\frac{\partial\phi^{(2)}}{\partial\xi}+l_{x}\frac{\partial u_{x}^{(2)}}{\partial\xi}\notag\\
&&+l_{y}\frac{\partial u_{y}^{(2)}}{\partial\xi}+\frac{l_{z}}{v_{0}}\frac{\partial u_{z}^{(1)}}{\partial\tau}. \label{abc}
\end{eqnarray}
Substituting the expressions for $n_{d}^{(2)}$, $u_{x}^{(2)}$ and $u_{y}^{(2)}$ from Eqs. \eqref{poisson-2}, \eqref{ux2} and \eqref{uy2} into Eq. \eqref{abc}, and noting that the coefficient of $\phi^{(2)}$ vanishes by the dispersion relation \eqref{disp-rel}, we obtain, after a few steps, the following KdV-like equation
\begin{equation}
\frac{\partial\phi^{(1)}}{\partial\tau}+A\sqrt{-\phi^{(1)}}\frac{\partial\phi^{(1)}}{\partial\xi}+B\frac{\partial^{3}\phi^{(1)}}{\partial\xi^{3}}=0. \label{kdv}
\end{equation}
where the coefficients of   nonlinearity and dispersion are
\begin{equation}
A=\frac{1-\sigma}{\sqrt{\pi}}\frac{\mu l^2_{z}}{v_0[\mu+\delta(1-\beta)]^2},
\end{equation}
\begin{eqnarray}
B&&=\frac{1}{6}v_{0}\frac{3+\left[\mu+\delta(1-\beta)\right]\left(5T-3v_{0}^{2}\right)}{\omega_{c}^{2}\left[\mu+\delta(1-\beta)\right]}\notag\\
&&+\frac{l^2_z}{2v_0\left[\mu+\delta(1-\beta)\right]^2},
\end{eqnarray}
in which $\mu=1\pm\delta$ for negatively/positively charged dusts.

 We note that in contrast to the usual KdV equation appeared in various contexts of plasmas (See, e.g., Refs. \cite{misra2013}),  Eq. \eqref{kdv}  has a nonlinear term proportional to the three-half power of the wave potential.  This modification of the  nonlinear term is, however, due to  the consideration of vortex-like ion distribution in the plasma. The similar form of the KdV equation has also been appeared     in unmagnetized dusty plasmas \cite{mamun1998b}.  We note that both the coefficients $A$ and $B$ are   modified by the parameters  $l_z$, $T$, $\delta$ and $\beta$,  while the effects of the trapped ions characterized by $\sigma$ and the static magnetic field characterized by $\omega_c$  are, respectively,  entered  in $A$ and $B$ only. It should be mentioned that the negative sign in $\sqrt{-\phi^{(1)}}$ in the nonlinear term of  Eq. \eqref{kdv}, which appears due to the vortex-like distribution of the positively charged particles, i.e.,  ions \cite{mamun1998b}    does not  appear in a recent work \cite{paul2013} on DA waves in an unmagnetized plasma with mobile dusts, nonthermal electrons and trapped positive ions. Furthermore,  in  Ref. \cite{paul2013} it has been shown that the DA soliton exists only of the compressive type with a form of sech$^2(\eta)$ as the KdV soliton. We, however,  show that the evolution equation for the propagation of DA solitary waves in dusty plasmas with trapped ions can be described by  a KdV-like equation  \eqref{kdv}, not in the form as derived in Ref.  \cite{paul2013}. Also,  as in the previous study \cite{mamun1998b},  Eq. \eqref{kdv} must have a stationary soliton in the form of \text{sech}$^4$-type (to be shown in the next section) not the \text{sech}$^2$-type.   We also show that unlike compressive solitons to exist in Ref.  \cite{paul2013}, the dusty plasmas with nonthermal fast electrons and trapped ions support only rarefactive-type solitons.

\section{Solution of the  KdV-like equation and discussion}

The travelling wave solution of Eq. \eqref{kdv} can be obtained  by transforming the independent variables $\xi$ and $\tau$ to $\eta=\xi-u_{0}\tau$ and $\tau=\tau$ where $u_{0}$ is a constant velocity normalized by $c_{d}$, and imposing the appropriate boundary conditions, namely, $\phi\rightarrow 0$, $d\phi^{(1)}/d\eta \rightarrow 0$, $d^{2}\phi^{(1)}/d\eta^{2}\rightarrow 0$ as $\eta\rightarrow\pm\infty.$
Thus, one can find  the soliton  solution  as (See for details, Appendix A)
\begin{equation}
\Phi\equiv\phi^{(1)}=-\phi_{m}^{(1)}\text{sech}^4\left(\frac{\xi-u_{0}\tau}{\Delta}\right), \label{soliton-sol}
\end{equation}
where the amplitude $\phi_{m}^{(1)}$ and the width $\Delta$ (normalized by $\lambda_{D}$) are given by
\begin{equation}
\begin{split}
&\phi_{m}^{(1)}=\left(\frac{15u_{0}}{8A}\right)^{2},\\
&\Delta=\left(\frac{16B}{u_{0}}\right)^{{1}/{2}}.
\end{split}
\end{equation}
This soliton solution   corresponds to the rarefaction or compression  of the   dust number density, i.e., the dust density holes or humps [See Eq. \eqref{poisson-1}] according to when  $\beta\gtrless2+1/\delta$.  From Eq. \eqref{soliton-sol} we find that in contrast to the usual KdV soliton (See, e.g., Ref. \cite{misra2013}),  the amplitude of the  soliton [Eq. \eqref{kdv}] does not depend on the sign of the nonlinear coefficient $A$. However, for the existence of such solitons, $B$ must be positive.  It can be shown that  $B>0$   for a wide range of values of the parameters, namely $0<l_z,T\lesssim1$, $\delta>1$  and $\beta>\beta_c$.

\begin{figure*}[ht]
\centering
\includegraphics[height=3in,width=6in]{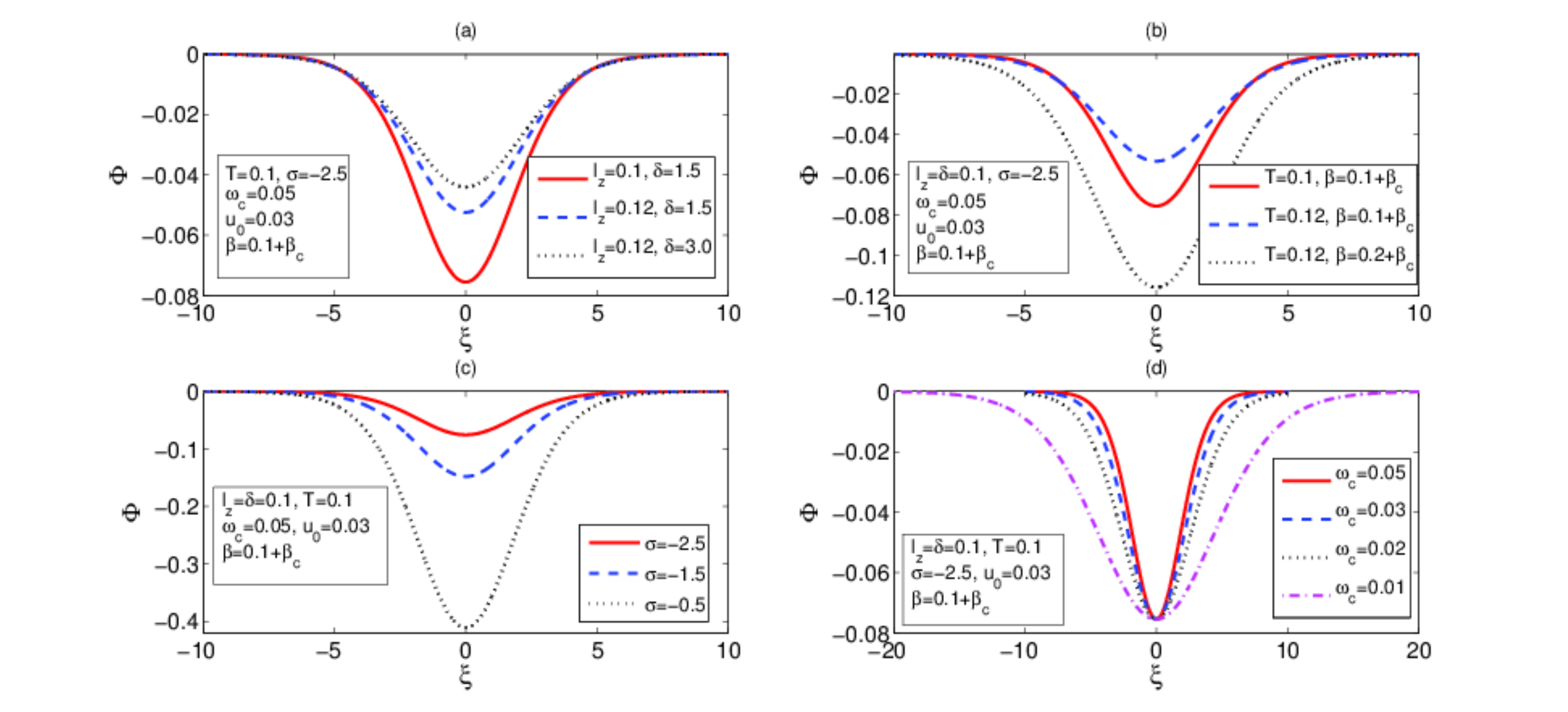}
\caption{Plot of the soliton solution [Eq. \eqref{soliton-sol} with $\xi$ at $\tau=0$ for different values of the parameters as in the figure. The effects of the static magnetic field on the width of the soliton remain unaltered for $\omega_c\gtrsim0.1$. }
\label{fig:figure2}
\end{figure*}

 It is pointed  out that  the perturbation method, which is  valid  for the small but finite amplitude perturbations may not be valid for large $\Phi$,  which makes the wave amplitude large enough to break the condition $\epsilon |\phi^{(1)}|<1$. We numerically analyze the properties of the stationary soliton [Eq. \eqref{soliton-sol}] by the effects of the parameters $l_z,~\delta,~T,~\beta,~\sigma$ and $\omega_c$. We consider those plasma parameters for which $\beta>\beta_c$ and $B>0$ are satisfied.  Different characteristic features of the DA soliton are exhibited in Fig. \ref{fig:figure2}.  Inspecting on the coefficients $A$ and $B$, we find that the external magnetic field characterized by $\omega_c$ (the parameter $\sigma$ for trapped ions) does  not have any effect on the amplitude (width), but on the width (amplitude) of the soliton. Next, we   study the effects of the plasma parameters as follows:

(i) \textit{Effects of the obliqueness of   propagation $(l_z)$ and the plasma density   $(\delta)$}: The effects of the obliqueness parameter $l_z$ and the electron to dust density ratio $\delta$ on the amplitude and width of the soliton are shown in Fig. \ref{fig:figure2}(a) (See the solid and  dashed lines for the variation with $l_z$, and dashed and dotted lines for $\delta$). It is found that, in each case, increasing the values of $l_z$ (which means that the cosine of the angle of propagation decreases) and $\delta$ (i.e., one enters into a relatively  dense plasma regime in which both electron and ion number densities increase in order to maintain the charge neutrality), keeping others constant  (i.e., $T=0.1,~\sigma=-2.5,\omega_c=0.05,~\beta=0.1+\beta_c$ and $u_0=0.03$), lead  to a decrease in the wave amplitude (Hereafter, the wave amplitude means its magnitude $|\Phi|$), and a small increase in the width.

(ii) \textit{Effects of the dust temperature $(T)$ and the presence of fast electrons $(\beta)$}:  The effects of the thermal pressure of charged dusts (solid and dashed lines) as well as the presence of fast electrons (See the dashed and dotted lines) in the plasma (Keeping others constant, i.e., $l_z=\delta=0.1,~\sigma=-2.5,\omega_c=0.05,~\beta=0.1+\beta_c$ and $u_0=0.03$),) on the soliton characteristics are exhibited in Fig. \ref{fig:figure2}(b).  It is seen that the temperature ratio $T$ of dust to electron temperature has stronger influence with decreasing   the amplitude and a small increase in the width of the soliton. The effect of $\beta$ is quite noticeable, i.e.,  a small increase in its value ($>\beta_c$) leads to a significant increase of both the amplitude and width of the soliton. Thus, in magnetized warm dusty nonthermal plasmas, the percentage of fast particles has to be larger than the slow counterparts in order to sustain the DA solitary waves, and higher the percentage of fast particles the larger are the   amplitude and width of the solitons.

(iii) \textit{Effects of the free to trapped ion temperature $(\sigma)$}: Figure  \ref{fig:figure2}(c) shows the effects of the presence of trapped ions characterized by $\sigma$ in the plasma. We find that when $\sigma<0$, which corresponds to a depression in the trapped particle distribution, its higher values (in magnitude) lead to a significant decrease of   the soliton amplitude. Since $B$ is independent of $\sigma$, the soliton width remains unchanged with the variation of $\sigma$.  On the other hand, when $\sigma~(>0)$ tends to $1$, i.e., as one approaches thermal equilibrium of ions in the plasma, the soliton amplitude is found to increase (not shown in the figure).

(iv) \textit{Effects of the static magnetic field $(\omega_c)$}: The influence of the magnetic field strength characterized by the nondimensional dust gyrofrequency $\omega_c$  on the properties of solitons is shown in Fig.  \ref{fig:figure2}(d). As mentioned before, the magnetic field does not have any effect on the amplitude of the soliton, however, it causes the solitons to become narrower with higher values of $\omega_c<0.1$. For  $\omega_c\gtrsim0.1$, the effect of the magnetic field on the soliton width is  no longer appreciable. The effect of $\omega_c$ is significant at its lower values, and thus, lower the values of $\omega_c$, the higher is the soliton width.

\begin{figure}[ht]
\centering
\includegraphics[height=1.5in,width=1.5in]{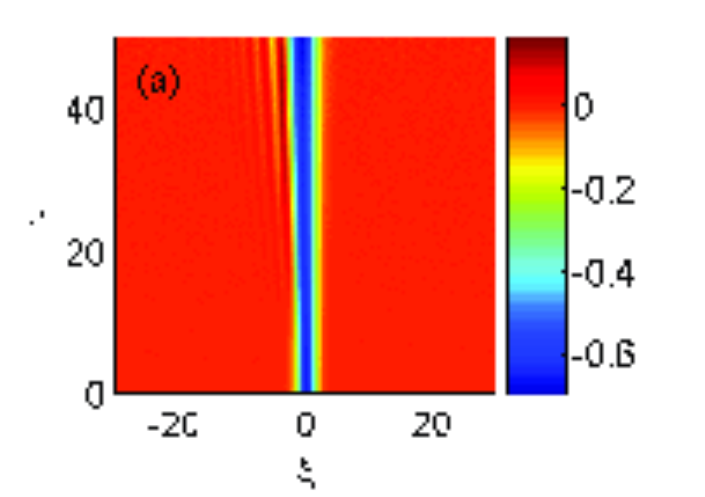}
\quad
\includegraphics[height=1.5in,width=1.5in]{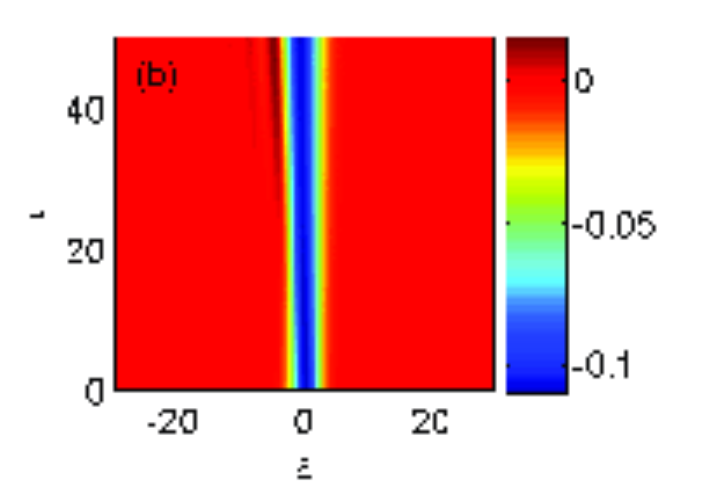}
\includegraphics[height=1.5in,width=1.5in]{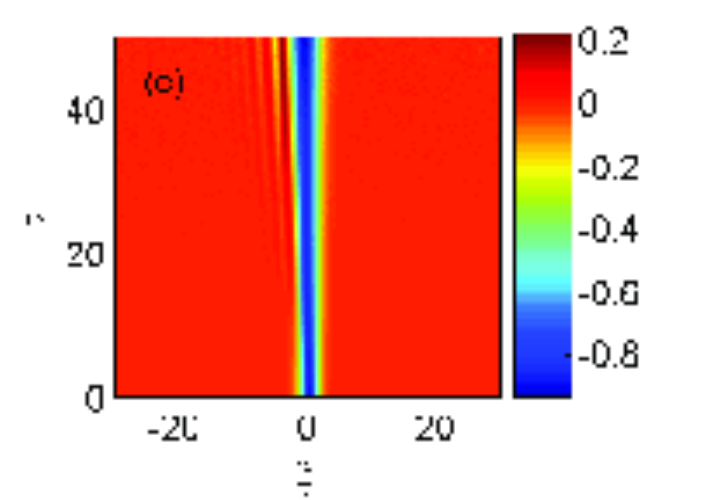}
\quad
\includegraphics[height=1.5in,width=1.5in]{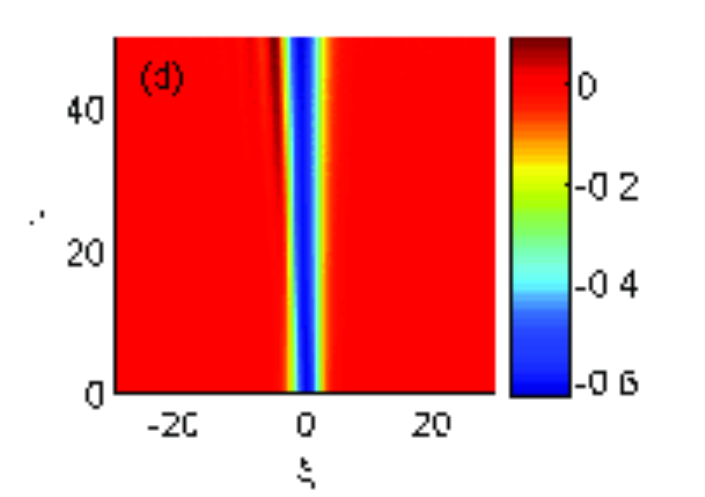}
\caption{The stability and instability of the DA solitons given by Eq. \eqref{kdv} are shown with the variations of the obliqueness parameter $(l_z)$ [See subplots (a) for $l_z=0.05$  and (b) for $l_z=0.12$ with a fixed $\sigma=-2.5$, $\beta=0.2+\beta_c$,  $\delta=1.5$, $T=0.1$  and $\omega_{c}=0.2$] and the free to trapped ion temperature $(\sigma)$ [See plots (c) for $\sigma=-2.0$, the other parameters being the same as in the subplot (a), and (d) for $\sigma=-0.5$, the other parameters being the same as in the subplot (b)].  }
\label{fig:figure3}
\end{figure}
\begin{figure}
\centering
\begin{minipage}[b]{0.5\textwidth}
\includegraphics[width=8.4cm]{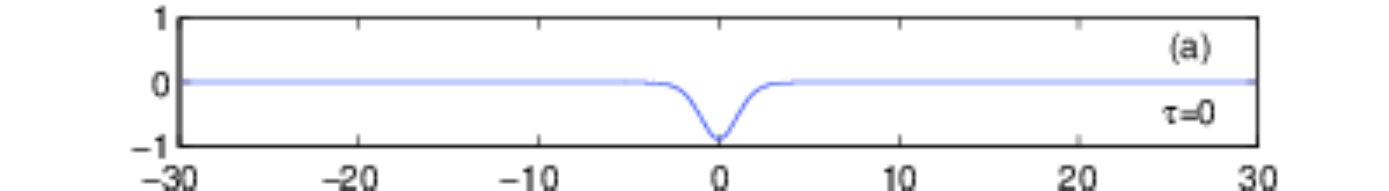}\\
\includegraphics[width=8.4cm]{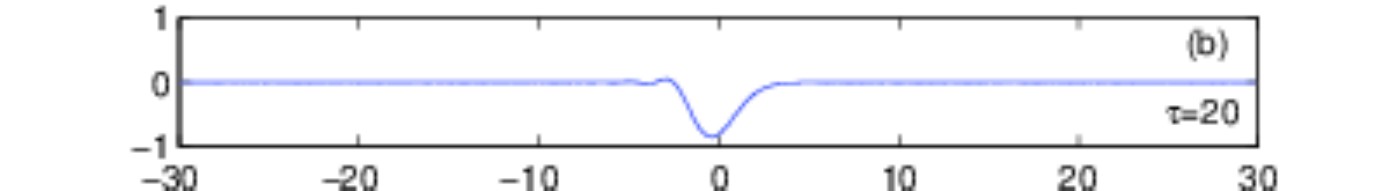}\\
\includegraphics[width=8.4cm]{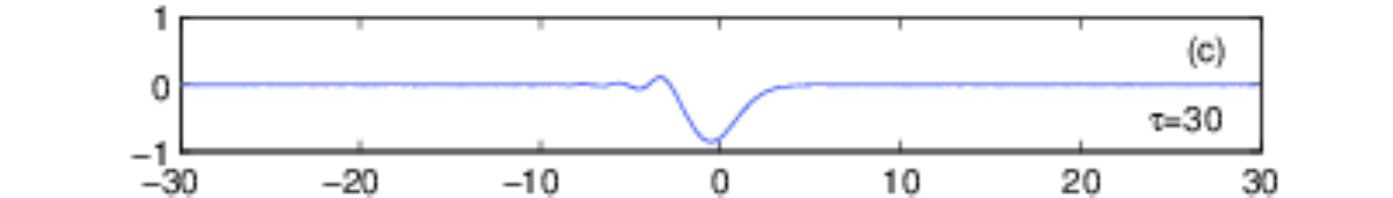}\\
\includegraphics[width=8.4cm]{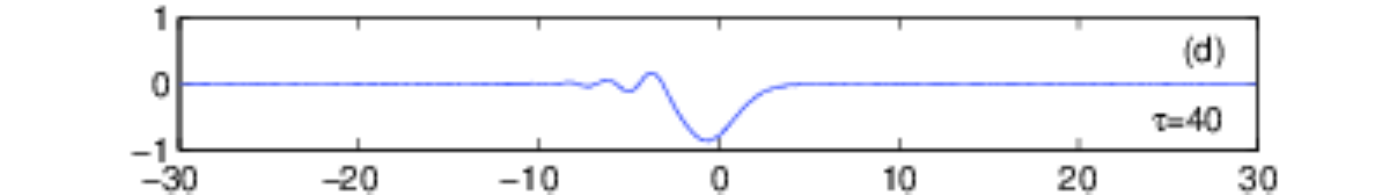}\\
\includegraphics[width=8.4cm]{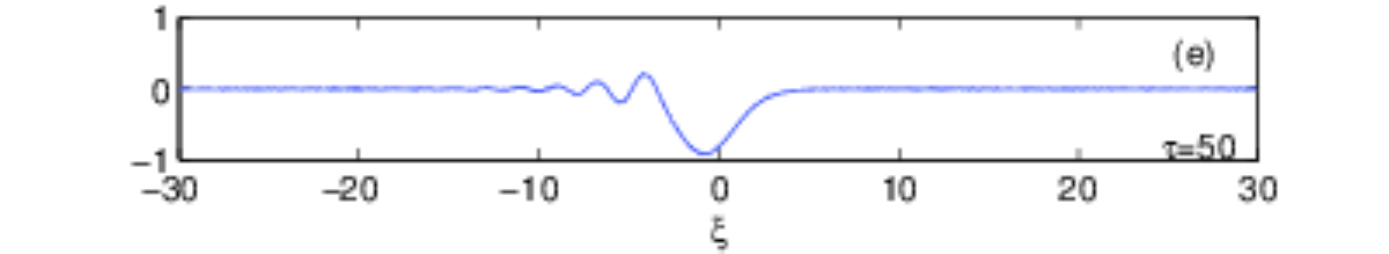}
\end{minipage}
\caption{Numerical solution of Eq. \eqref{kdv} exhibiting  soliton profiles   at different times: (a) $\tau=0$, (b) $\tau=20$, (c) $\tau=30$, (d) $\tau=40$, and (e) $\tau=50$. The parameters values are the same as those in Fig. \ref{fig:figure3}(c).}
\label{fig:figure4}
\end{figure}

We numerically investigate the time evolution as well as the stability/instability of the DA solitary waves given by Eq. \eqref{kdv}.  For the numerical integration of Eq. \eqref{kdv}, we use the centered second-order difference approximations for the spatial derivatives with periodic boundary conditions and the standard fourth-order Runge-Kutta scheme for the time stepping. The simulation  results are exhibited in Fig. \ref{fig:figure3}. The effects of the oblique wave propagation $(l_z)$ on the stability of the DA solitary waves  are shown in Figs. 3(a)   and 3(b)   for    $l_z=0.05$ and   $l_z=0.12$ respectively. The other fixed parameters are $\sigma=-2.5$, $\beta=0.2+\beta_c$,  $\delta=1.5$, $T=0.1$,   and $\omega_{c}=0.2$.  It is found that the solitary waves are more stable for larger $l_z<1$, i.e., as the angle of propagation to the static magnetic field decreases, the stability of the DA perturbations increases. Figures 3(c)  and 3(d)   show the similar effects of the trapped ion temperature (characterized by $\sigma$) on the solitary waves
 for  $\sigma=-2.0$, $l_z=0.05$ and   $\sigma=-0.5$ $l_z=0.12$ respectively. The other fixed parameters are the same as   those used for Figs. 3(a)  and 3(b).
Comparing Figs. 3(a)  and 3(c), and    Figs. 3(b)   and 3(d), we find that as   $|\sigma|$ ($\sigma<0$) decreases, i.e., when the magnitude of the trapped ion temperature is higher than that of the free ions,  the instability of the waves increases.  The initial conditions used for the plots of Figs. 3(a)-3(d), respectively, are   
(i) $\phi ^ {(1)}=-0.065~ \text{sech}^4(\xi /2.002)$,
(ii) $\phi^ {(1)}=-0.113~ \text{sech}^4(\xi /3.099)$,
(iii) $\phi^ {(1)}=-0.884~ \text{sech}^4(\xi /2.002)$, and
(iv) $\phi^ {(1)}=-0.614~ \text{sech}^4(\xi /3.099)$. These are the steady-state solutions of Eq. \eqref{kdv} at $\tau=0$. From Fig. \ref{fig:figure3}, we also find that the solitons propagate in the negative $\xi$-direction with a profile (almost unchanged) for $\tau<20$. However, as time goes on, an oscillatory structure is seen to grow  in  front of the  soliton.

It is to be noted that the other parameters, namely $\delta$, $T$, and $\omega_c$ can also have the similar influence on the stability/instability of the solitons. It is found that an increase of $T$   improves the stability of the soliton, while an increase of $\delta$  enhances the instability of the soliton. However, the instability of the soliton is not greatly affected by the variation of $\omega_c$. Since the values of the nonthermal parameter $\beta$ $(=0.2+\beta_c)$ depends on $\beta_c=1+(1+\delta+3/5T)/\delta$, one can also see the effects of $\beta$ with the variations of $T$ and $\delta$ as above.

The dynamical evolution of the soliton profiles is exhibited in Fig.\ref{fig:figure4}  at different times: (a) $\tau=0$, (b) $\tau=20$, (c) $\tau=30$, (d) $\tau=40$, and (e) $\tau=50$. The parameters and the initial condition are the same as those in Fig.  3(c). It is seen that in contrast to the KdV soliton (in which the wave steepening occurs \cite{misra2012}), the leading part of the initial pulse flattens due to the modified nonlinearity (proportional to the three-half power of the wave potential). As the time progresses, the pulse separates into solitons that are down-shifted and a residue due to the wave dispersion.  It is clear from Fig. \ref{fig:figure4}(d) that once the solitons are formed and get separated, they propagate almost without changing their shape due to the balance of the nonlinearity and the wave dispersion.  From Figs.  \ref{fig:figure3} and \ref{fig:figure4}, one can conclude that the DA solitons can withstand perturbations and turbulence in a finite interval of time.  It is to be mentioned that the time evolution of the KdV equation \eqref{kdv} as well as the stability/instability of its solitary solution can also be studied by some other methods, e.g., semi-analytical methods \cite{hajmohammadi2014,hajmohammadi2012,amour2011}.

\section{Conclusion}

We have investigated the propagation characteristics of electrostatic  dust-acoustic  waves in a magnetized dusty plasma with the combined effects of the nonthermal electrons and  trapped ions with vortex-like distribution.  By using the reductive perturbation technique, a KdV-like equation with a nonlinearity proportional to three-half power of the wave potential  is derived to investigate the nonlinear propagation of DA solitary waves. It is found that the  DA wave propagation is possible when the percentage of energetic electrons remains higher than the slow counterparts, i.e., the degree of nonthermality $\beta>1$ and   exceeds its critical value $\beta_c$. The latter changes with the number densities of the particles as well as the thermal pressure of charged dusts. In absence of the dust thermal pressure, i.e., in cold dusty  magnetoplasmas, DA solitary wave can propagate  with $\beta<1$ \cite{paul2013}.

The stationary soliton solution of the KdV-like equation, in the form of \text{sech}$^4(\eta)$ (Quite distinctive from the  KdV soliton), is obtained and its properties are studied with the system parameters. It is shown that  the external magnetic field characterized by $\omega_c$ (the parameter $\sigma$ for trapped ions) does  not have any influence on the amplitude (width), but on the width (amplitude) of the soliton. Thus,  the external magnetic field causes the solitary structures to become more spiky, while the soliton amplitude decreases (increases)  with $\sigma<0$ $(0<\sigma<1)$.
Furthermore, the dynamical evolution of the DA solitary waves shows
that the DA soliton can   withstand perturbations and turbulence during a considerable time. Our numerical results show  that a decrease of the obliqueness parameter $l_z$, the ratio of dust to electron temperature $T$,  as well as the ratio of the free to trapped ion temperature $|\sigma|$ ($\sigma<0$) favors the instability of the DA solitons, while the similar instability can be achieved by an increase of the ratio of electron  to dust charge density $(\delta)$. 
 
 To conclude,  the theoretical results should be useful for understanding the nonlinear features of electrostatic dust-acoustic solitary waves that propagate obliquely to the magnetic field in nonthermal plasmas with higher percentage of energetic (fast) electrons and trapped ions in laboratories as well as space plasmas (e.g., Earth's magnetosphere, auroral region, heliospheric environments etc.).

\section*{Acknowledgment}
{This research was partially supported by the SAP-DRS (Phase-II), UGC, New Delhi, through sanction letter No. F.510/4/DRS/2009 (SAP-I) dated 13 Oct., 2009, and by the Visva-Bharati University, Santiniketan-731 235, through Memo No. Aca-R-6.12/921/2011-2012 dated 14 Feb., 2012. Y.W.  acknowledgs support from NSFC (No. 11104012) and the Fundamental Research Funds for the Central Universities (Nos. FRF-TP-09-019A, FRF-BR-11-031B).}

\appendix 
\section{ Stationary soliton solution of the KdV-like equation}
Equation \eqref{kdv}  can be rewritten as
\begin{equation}
\frac{\partial\Phi}{\partial\tau}+iA\sqrt{\Phi}\frac{\partial\Phi}{\partial\xi}+B\frac{\partial^{3}\Phi}{\partial\xi^{3}}=0. \label{eq0}
\end{equation}
From Eq. \eqref{eq0}, using the transformation $\eta=\xi-u_0\tau$,  we get
\begin{equation}
\frac{d}{d\eta}\left(B \ddot{\Phi}-u_0\Phi+i\frac{2}{3}A\Phi^{3/2}\right)=0, \label{eq1}
\end{equation}
where the dot denotes differentiation with respect to $\eta$. Integrating Eq. \eqref{eq1} and using the boundary conditions $\Phi,~\ddot{\Phi}\rightarrow0$ as $\xi\rightarrow\pm\infty$ we get
\begin{equation}
B \ddot{\Phi}-u_0\Phi+i\frac{2}{3}A\Phi^{3/2}=0. \label{eq2}
\end{equation}
Multiplying Eq. \eqref{eq2} by $2\dot{\Phi}$ and integrating with respect to $\eta$, we obtain
 \begin{equation}
B \dot{\Phi}^2-u_0\Phi^2+i\frac{8}{15}A\Phi^{5/2}=0, \label{eq3}
\end{equation}
where we have used the boundary conditions $\Phi,~\dot{\Phi}\rightarrow0$. From Eq. \eqref{eq3} we have
\begin{equation}
\dot{\Phi}=\pm\Phi\sqrt{\frac{u_0}{B}-i\frac{8A}{15B}\sqrt{\Phi}},
\end{equation}
\begin{equation}
\text{or,~}\int\frac{d\Phi}{\Phi\sqrt{{u_0}/{B}-i\left({8A}/{15B}\right)\sqrt{\Phi}}}=\pm \int d\eta,\label{eq4}
\end{equation}
which gives ($a=u_0/B$ and $b=i8A/15B$)
\begin{equation}
 -\frac{4}{\sqrt{a}}\tanh^{-1}\sqrt{\frac{a-b\sqrt{\Phi}}{a}}=\pm\eta,
 \end{equation}
 \begin{equation}
 \text{or,~}\sqrt{\frac{a-b\sqrt{\Phi}}{a}}=\mp\tanh\left(\frac{\sqrt{a}}{4}\eta\right).\label{eq5}
\end{equation}
Thus, we obtain the soliton solution as
\begin{equation}
1-\tanh^2\left(\frac{\sqrt{a}}{4}\eta\right)= \frac{b}{a}\sqrt{\Phi},
\end{equation}
\begin{equation}
\text{or,~}\Phi=-\left(\frac{15u_0}{8A}\right)^2\text{sech}^4\left(\sqrt{\frac{u_0}{16B}}\eta\right).\label{eq6}
\end{equation}

\bibliographystyle{elsarticle-num}

\end{document}